\documentclass[aps,prb,reprint,superscriptaddress,noeprint]{revtex4-2}
\usepackage{siunitx}
\usepackage{graphicx}			

\usepackage{color}
\usepackage{orcidlink}

\begin{document}
\newcommand{\bsts}{BiSbTeSe$_2$}     
\newcommand{\bsb}{Bi$_{1.5}$Sb$_{0.5}$Te$_{1.7}$Se$_{1.3}$}
\newcommand{\bsn}{Bi$_{1.08}$Sn$_{0.02}$Sb$_{0.9}$Te$_{2}$S}

\newcommand{\tfrac}[2]{{\footnotesize \frac{#1}{#2}}} 
\newcommand{\boldsymbol}[1]{\textbf{\textit{#1}}} 

\title{Phonon thermal Hall effect in charge-compensated topological insulators}

\author{Rohit Sharma\,\orcidlink{0000-0001-9815-0733}}
\email{sharma@ph2.uni-koeln.de}
\affiliation{II.\, Physikalisches Institut, Universit\"at zu K\"oln, Z\"ulpicher Str.\ 77, 50937 K\"oln, Germany}
\author{Mahasweta Bagchi\,\orcidlink{0000-0002-0031-0952}}
\affiliation{II.\, Physikalisches Institut, Universit\"at zu K\"oln, Z\"ulpicher Str.\ 77, 50937 K\"oln, Germany}
\author{Yongjian Wang\,\orcidlink{0000-0001-5921-1986}}
\affiliation{II.\, Physikalisches Institut, Universit\"at zu K\"oln, Z\"ulpicher Str.\ 77, 50937 K\"oln, Germany}
\author{Yoichi Ando\,\orcidlink{0000-0002-3553-3355}}
\affiliation{II.\, Physikalisches Institut, Universit\"at zu K\"oln, Z\"ulpicher Str.\ 77, 50937 K\"oln, Germany}
\author{Thomas Lorenz\,\orcidlink{0000-0003-4832-5157}}
\email{tl@ph2.uni-koeln.de}
\affiliation{II.\, Physikalisches Institut, Universit\"at zu K\"oln, Z\"ulpicher Str.\ 77, 50937 K\"oln, Germany}

\date{\today}

\begin{abstract}

 From a systematic study of thermal and charge transport in various single crystals of compensated topological insulators we identify the evolution of a large low-temperature thermal Hall effect as a characteristic common feature. In order to separate phononic and electronic contributions in the measured longitudinal and transverse thermal conductivity, the electronic contributions are estimated from corresponding electrical resistivity and Hall effect measurements on the same samples by using the Wiedemann-Franz law. As may be expected  for charge-compensated topological insulators the longitudinal thermal conductivity is phonon-dominated in all samples. However, we also find a pronounced field-linear thermal Hall effect that becomes most pronounced in the low-temperature range, where all samples are good electrical insulators. 
This indicates an underlying phononic mechanism of the thermal Hall effect and in this respect the topological insulators resemble other, mainly ionic, insulators, which have been reported to show a phonon-induced thermal Hall effect, but its underlying phononic mechanism remains to be identified. Our observation of a comparable thermal Hall ratio in topological insulators supports a theoretical scenario that explains a thermal Hall effect through skew scattering on charged impurities.
\end{abstract}


\maketitle

\section{Introduction}

The thermal Hall effect (THE) has gained significant scientific interest due to its unique ability to explore various excitations, like electrons, phonons, magnons and more exotic excitations like Majorana particles\cite{kasahara2018majorana}. This sets it apart from the charge Hall effect, which only provides insights into mobile charged excitations, such as electrons. THE is the transverse temperature difference ($\Delta{T}_y$) that emerges along the $y$ axis when a heat current ($J$) and a magnetic field ($H$) are applied along the $x$ and $z$ axes, respectively. If charged particles contribute to the heat transport a finite THE is naturally expected from the Lorentz force and the corresponding heat and charge conductivities can be related via the Wiedemann-Franz law.
However, despite the absence of charged carriers, insulators can also exhibit a THE as was first observed in paramagnetic Tb$_3$Ga$_5$O$_{12}$~\cite{PhysRevLett.95.155901}. This observation prompted theoretical discussions considering Raman-type interactions between large spins and phonons, the Berry curvature of phonon bands, and skew scattering of phonons by superstoichiometric Tb$^{3+}$ ions \cite{PhysRevLett.96.155901, PhysRevB.86.104305, PhysRevLett.113.265901}.
Another source of THE in magnetic insulators is associated with the Berry curvature of magnon bands generated by the Dzyaloshinskii-Moriya interaction~\cite{PhysRevB.85.134411,2010Sci...329..297O,PhysRevLett.115.106603,2021PhRvL.127x7202Z}.

Subsequently, phonon-induced THEs have been detected in various materials, such as multiferroic materials like Fe$_2$Mo$_3$O$_8$ \cite{2017NatMa..16..797I}, cuprate superconductors \cite{2019Natur.571..376G, 2020NatPh..16.1108G}, phonon glass systems like Ba$_3$CuSb$_2$O$_9$ \cite{PhysRevLett.118.145902}, Mott insulators \cite{2020NatCo..11.5325B}, the metallic spin-ice material Pr$_2$Ir$_2$O$_7$ \cite{2022NatCo..13.4604U}, and the antiferromagnetic insulator Cu$_3$TeO$_6$ \cite{2022PNAS..11908016C}.
Remarkably, even non-magnetic SrTiO$_3$ exhibits a substantial phonon-induced THE that has been attributed to the presence of antiferrodistortive structural domains \cite{PhysRevLett.124.105901}, and this THE is suppressed in isotopically substituted SrTi$^{18}$O$_3$ \cite{PhysRevLett.126.015901} and Ca-doped samples \cite{2022PNAS..11901975J}. A phonon-induced THE has been identified in elemental black phosphorus \cite{2023NatCo..14.1027L}, another nonmagnetic material, where the coupling of phonons with magnetic fields has been connected to anisotropic charge distribution. Theoretically two proposals for phonon induced THE have been put forward. On the one hand, intrinsic mechanisms consider a phonon THE arising from factors like Berry curvature in phonon bands \cite{PhysRevB.86.104305}, phonon scattering due to collective fluctuations \cite{PhysRevB.106.245139}, or interactions between phonons and other quasi-particles like magnons~\cite{{PhysRevB.104.035103},{PhysRevLett.123.167202}}. On the other hand, extrinsic mechanisms rely on effects of phonon scattering by impurities or defects \cite{{2022PNAS..11915141G},{PhysRevB.103.205115},{PhysRevB.105.L220301}}.
Despite intensive research, the exact origins of phonon-induced THE remain elusive. This highlights the importance of exploring various material classes, especially beyond oxide-based ionic insulators.

\begin{figure*}[tbp]
	\centering
	\hspace*{-0.5cm}
	\includegraphics[width=\textwidth]{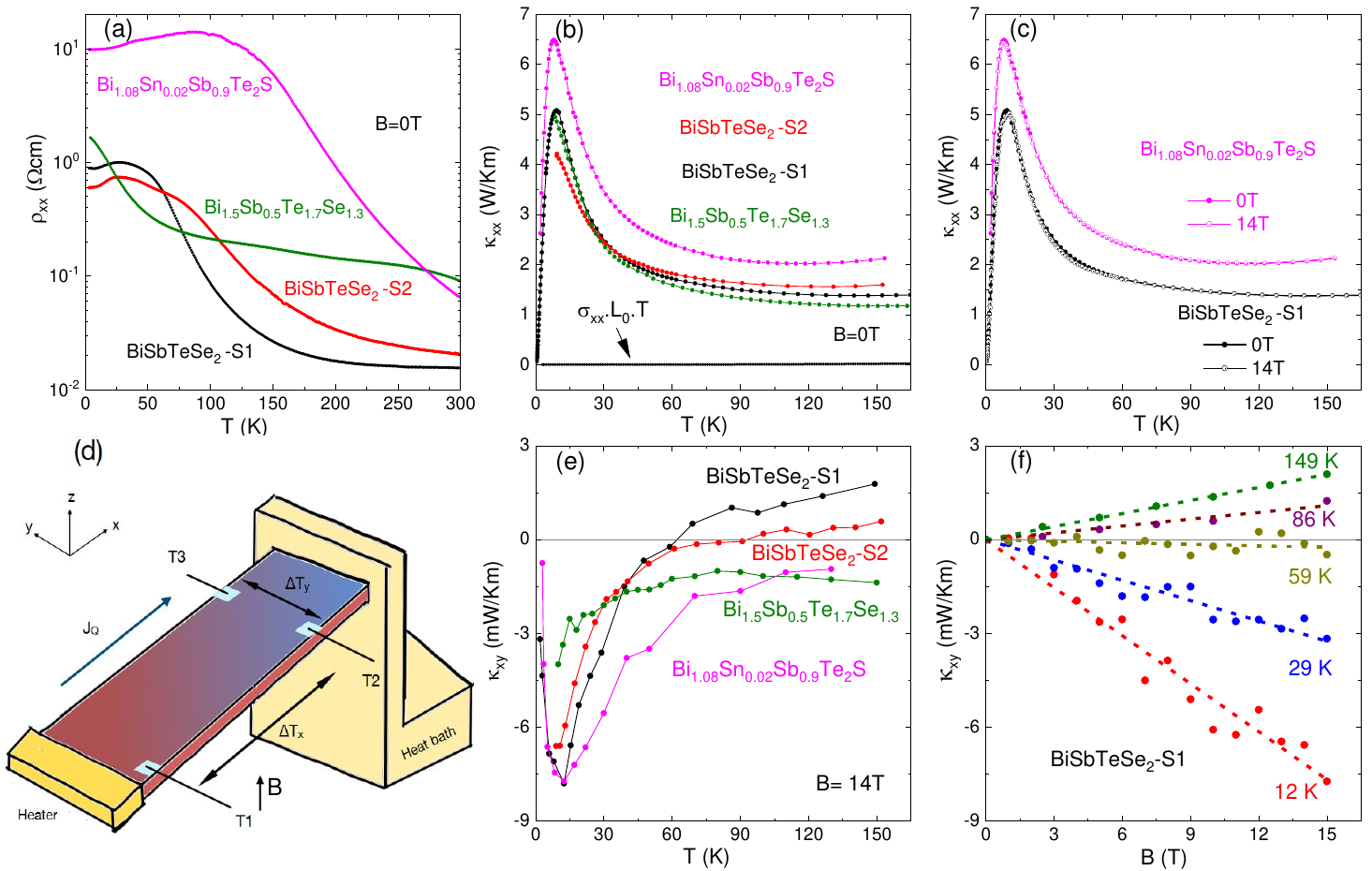}
	\caption{Temperature dependence of (a) electrical resistivity $\rho_{xx}$ and of (b) thermal conductivity $\kappa_{xx}$ of various charge compensated topological insulators. The flat black line close to zero in (b) represents a negligible charge-carrier contribution $\kappa_{xx}^{el}=\sigma_{xx}L_0T$ estimated via the Wiedemann-Franz law for the sample \bsts -S1 with the lowest resistivity $\rho_{xx}=1/\sigma_{xx}$. This reveals that $\kappa_{xx}$ is phonon dominated for all samples and is further confirmed in (c) showing the magnetic-field independence of $\kappa_{xx}$. (d) Schematic setup to measure $\kappa_{xx}$ and $\kappa_{xy}$ from the longitudinal and transverse temperature differences ($\Delta T_x = T1-T2$, $\Delta T_y = T2-T3$) induced by a heat current $J_Q \|x$ in a magnetic field $H\,| z$. (e) Temperature-dependent $\kappa_{xy}$ in a magnetic field $\mu_0 H = 14\,$T. For each sample, $\kappa_{xy}(T)$ was obtained from linear fits of field-antisymmetrized data measured at constant temperatures as is exemplarily shown in (f).}
	\label{FIG.1.}
\end{figure*}

Here, we report the discovery of a large phonon-related THE in a series of Bi$_{2-x}$Sb$_x$Te$_{3-y}$Se$_y$ samples that belong to a different class of non-magnetic, insulating materials known as compensated three dimensional (3D) topological insulators (TIs).  3D TIs are characterized by an insulating bulk band structure in combination with gapless (conducting) surface states that arise from  a non-trivial topology of the underlying wave functions~\cite{2013JPSJ...82j2001A}. Often, however, real TI materials are rather conducting due to defect-induced charge carriers and bulk insulating behavior is only reached by a specific compensation of donor and acceptor atoms~\cite{PhysRevB.84.165311}.  From thermal and electrical transport measurements we identify a large THE as a common feature of several bulk insulating TIs. In a field of 14~T, large thermal Hall ratios $\kappa_{xy}/\kappa_{xx} \simeq -10^{-3}$ are reached and hardly vary over the low-temperature insulating phases. This compares well to the results obtained on many of the above mentioned insulators, see~\cite{2022PNAS..11908016C}, and is in line with a recent theoretical proposal of a phonon THE arising from charged defects in (nonmagnetic) insulators~\cite{PhysRevB.105.L220301}.

\section{Experimental}

Single crystals of compensated TIs were grown from a melt of high-purity elements Bi, Sb, Te, Se, S, and Sn, using a modified Bridgman method in a sealed quartz-glass tube, as described in~\cite{PhysRevB.84.165311}. Freshly cleaved thin flakes were cut to rectangular shape to measure the longitudinal ($\rho_{xx}$) and Hall resistivities ($\rho_{yx}$) in a standard 6-point geometry using a DC current source and two nanovoltmeters at stabilized temperatures and discrete magnetic fields between $\pm 14\,$T. The same samples were used for measuring the longitudinal ($\kappa_{xx}$) and Hall ($\kappa_{xy}$) thermal conductivities using a steady-state 1-heater 3-thermometer method under high-vacuum conditions, as sketched in Fig.~\ref{FIG.1.}. A 10~k$\Omega$ RuO$_2$ chip resistor at one end of the sample was used to induce a heat current $J_Q =RI^2$ that results in longitudinal $\Delta T_x=T_1-T_2$ and transverse $\Delta T_y=T_2-T_3$ temperature differences. Up to about 100~K, $T_{1,2,3}$ were measured with Cernox sensors (CX1030) and, additionally, $\Delta T_x$ and $\Delta T_y$ were measured with constantan-chromel thermocouples in separate runs from about 30 to 200~K. Gold wires were  contacted to the sample with silver paste and were either connected to a commercial PPMS puck (Quantum Design) for the electrical measurements, or connected to the respective temperature sensors of different home-built setups for heat transport  measurements. In order to eliminate the misalignment of the transverse contacts, the Hall voltage $U_H$ or temperature difference $\Delta T_H$ were obtained by antisymmetrization of the respective raw data measured in $\pm B$, e.g.\ $U_H(B)=(U_y(+B)-U_y(-B))/2$. The  thermal Hall conductivity is then obtained as $\kappa_{xy} = (\Delta T_H / \Delta T_x)\,(l / w)\,\kappa_{xx}$ with longitudinal thermal conductivity $\kappa_{xx} = (J_Q / \Delta T_x)(l / wt)$, distance $l$ between the longitudinal contacts, sample width $w$, and thickness $t$. Analogously, the electrical resistivities are given by $\rho_{xx} = (U_{xx} / I_x)(wt /l)$ and $\rho_{yx} = (U_{H} t/ I_x)$, and the electrical Hall conductivity is $\sigma_{xy}=\rho_{yx}/(\rho_{xx}^2+\rho_{yx}^2)$. 

\section{Results and discussion}

Figure~\ref{FIG.1.}(a) shows the longitudinal electrical resistivity of four TI samples. Two of them have the same composition BiSbTeSe$_2$ and show a very similar $\rho_{xx}(T)$ characterized by a temperature-activated behavior above 100~K that turns into a low-temperature plateau. This is typical for these types of charge-compensated TIs and has been analyzed in detail previously~\cite{PhysRevB.84.165311}. Arrhenius fits  $\rho_{xx} \propto \exp\left(\Delta / k_{B}T\right)$ yield activation energies around 35~meV in good agreement with the previous results, whereas a larger activation energy of 136~meV and an overall enhanced resistivity  
is found in \bsn\ reflecting the higher degree of compensation in this type of TI material~\cite{2016NatCo...711456K}. In contrast, neither the activated behavior nor the low-temperature plateau are well pronounced in $\rho_{xx}(T)$ of \bsb , indicating more disorder and a weaker degree of charge-carrier compensation in this sample. Table~\ref{Table.1} in the Appendix compares the charge-carrier densities derived from the low-field Hall resistivity data of these samples for some characteristic temperatures.

The longitudinal thermal conductivity $\kappa_{xx}$ varies little between the four samples, as is shown in Fig.~\ref{FIG.1.}(b). The peaks around 10~K are typical for phonon heat transport and result from the opposite temperature dependencies of phonon heat capacity and phonon mean-free path which, respectively, increase and decrease with increasing temperature.   
In the related mother compounds Bi$_2$Se$_3$ and Bi$_2$Te$_3$ significantly higher $\kappa_{xx}$ peaks of $\sim$$\,35\,$W/Km are observed~\cite{navratil2004conduction,Yao2017}, whereas Sb$_2$Te$_3$ has a peak of $\sim$\,20\,W/Km that decreases to 6--$8\,$W/Km  when Sb is replaced by a few \% of V~\cite{Dyck2002}. The strong variation of the peak values is naturally attributed to the maximum phonon mean free path that sensitively depends on disorder effects, which vary with the composition and generally increase with the number of different constituents. Above $100\,$K, the $\kappa_{xx}$ of our samples are only very weakly temperature dependent and lie in the range between 1--$2\,$W/Km, which compares well with the values that are reached above about $200\,$K in the binary mother compounds~\cite{navratil2004conduction,Yao2017,Dyck2002}. 

Apart from the phonons, mobile charge carriers also contribute to the overall heat transport. The electronic contribution $\kappa_{xx}^{el}$ to the heat transport can be estimated from the measured electrical conductivity $\sigma_{xx}$ by using the Wiedemann-Franz law $\kappa_{xx}^{el} = L_0\,\sigma_{xx}\,T$ with temperature $T$ and Lorenz number L$_0 \approx 2.44\times10^{-8}$ V$^2$K$^{-2}$. In Fig.~\ref{FIG.1.}(b), the obtained electronic $\kappa_{xx}^{el}$ is exemplarily included for the BiSbTeSe$_2$-S1, but even for this best conducting sample $\kappa_{xx}^{el}$ remains 2 orders of magnitude below the  measured $\kappa_{xx}$ over the entire temperature range. Thus, we can safely conclude that the thermal conductivity in all our samples is phonon dominated and this is further verified by the fact that $\kappa_{xx}$ varies by less than 2\,\% in a magnetic field of 14~T in all samples. This is shown for two samples in Fig.~\ref{FIG.1.}(c) and more data of the weak field dependence of $\kappa_{xx}$ are presented in the Appendix, see Figs.~\ref{FIG.app_kxx_kxy}(a--d).

\begin{figure}[t]
	\centering
	\hspace*{-0.6cm}
	\includegraphics[width=0.98\columnwidth]{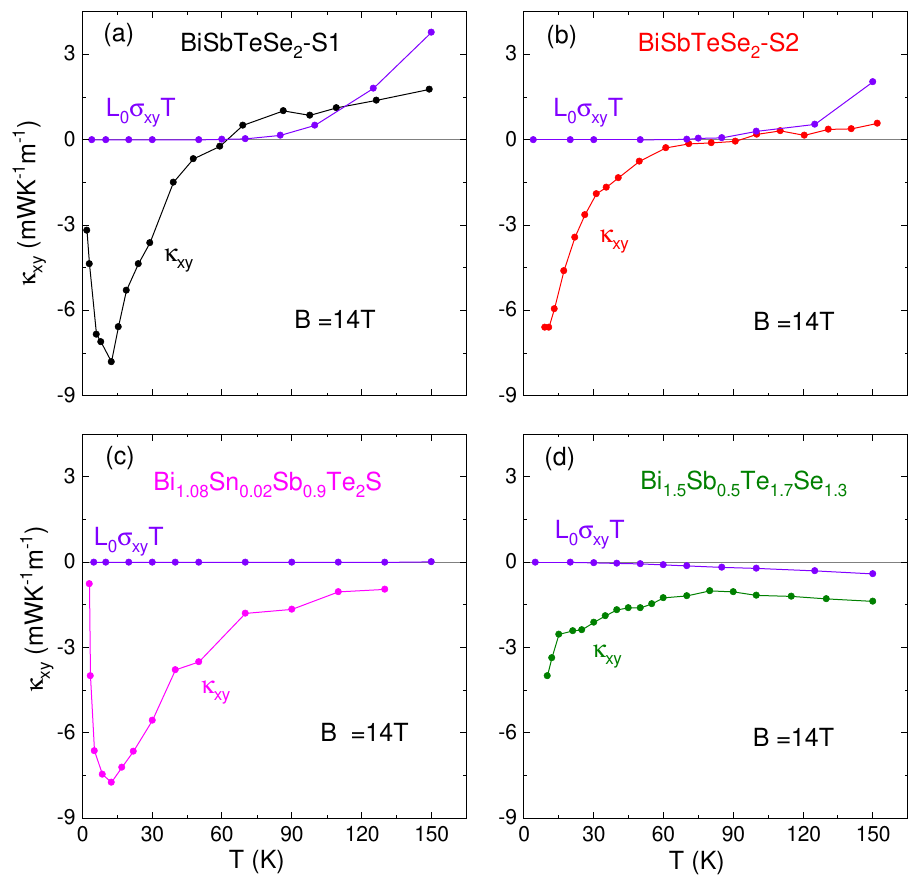}
	\caption{Temperature dependent $\kappa_{xy}$ data from Fig.~\ref{FIG.1.}(e) in comparison to the charge-carrier contributions $\kappa_{xy}^{el}=\sigma_{xy}L_0T$ estimated via the Wiedemann-Franz law from the electrical Hall conductivities $\sigma_{xy}$ measured in 14\,T. (a,b) $\kappa_{xy}$ of both \bsts\ samples change to a positive sign and roughly merge with the corresponding $\sigma_{xy}L_0T$ above $\sim\,$100\,K. (c) For \bsn , $\sigma_{xy}L_0T$ remains below 1\,\% of $\kappa_{xy}$, whereas (d) the curves of \bsb\ show comparable temperature dependences above $\sim\,$100\,K.}
	\label{FIG.2.}
\end{figure}

Having established that $\kappa_{xx}$ is fully phonon dominated, we direct our attention to $\kappa_{xy}$, which is sizable and shows a linear magnetic-field dependence  $\kappa_{xy}\propto B$ at constant $T$ for all samples over the entire temperature range. Representative $\kappa_{xy}(B,T)$ curves are displayed in Fig.~\ref{FIG.1.}(f) for one sample and the other data sets are shown in the Appendix. From linear fits of the $\kappa_{xy}(B)$ curves, we derive the temperature dependent $\kappa_{xy}(T)$ in a field of 14\,T that is compared for all four samples in Fig.~\ref{FIG.1.}(e). At higher temperature, the sign of $\kappa_{xy}$ is sample dependent, whereas  low-temperature peaks of negative sign evolve in all four samples with maximum values around $-6\,$mW/Km. In analogy to $\kappa_{xx}$, we estimate the expected electronic contribution to the thermal Hall conductivity by using the Wiedemann-Franz law $\kappa_{xy}^{el} = L_0\,\sigma_{xy}\,T$ with the electrical Hall conductivities $\sigma_{xy}$ measured on the same samples. As shown in Fig.~\ref{FIG.2.}, the positive $\kappa_{xy}$ at higher temperature in both \bsts\ samples coincide with hole-dominated electrical Hall conductivities and the estimated $\kappa_{xy}^{el}$ are of similar magnitude as the measured $\kappa_{xy}$. In \bsb , $\sigma_{xy}$ is electron-like at all temperatures and the obtained $\kappa_{xy}^{el}$ remains significantly smaller than $\kappa_{xy}$, but the temperature dependencies of both quantities become quite similar above $\sim$\,100\,K, whereas the obtained $\kappa_{xy}^{el}$ for \bsn\ remains below 1\,\% of $\kappa_{xy}$ up to the highest temperature. Thus we conclude that toward higher temperature thermally excited mobile charge carriers contribute to the THE in samples with a comparatively low excitation gap and a low resistivity. This contribution is weaker in \bsb\ with an overall enhanced resistivity signaling a reduced mobility of charge carriers, and becomes negligible in \bsn , which is the best-compensated TI with a very large excitation gap. The different behavior of mobile charge carriers in these samples is also reflected in different magnetic-field dependencies of the electrical Hall effect, which in all samples becomes non-linear below about 100\,K, see Appendix, but these differences are not in the focus of this report.

\begin{figure}[t]
	\centering
	\includegraphics[width=0.98\columnwidth]{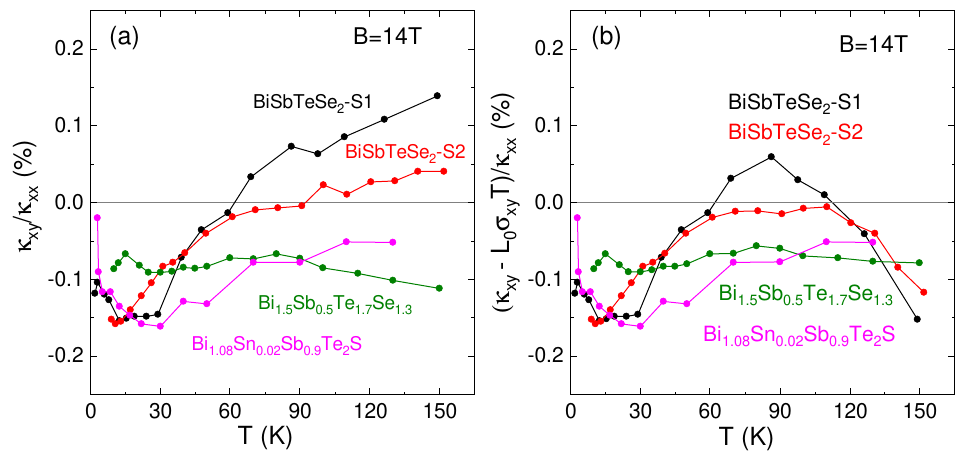}
	\caption{Temperature dependent thermal Hall ratio obtained by using either (a) the total $\kappa_{xy}$ or (b) the  estimated phonon Hall thermal conductivity $\kappa_{xy}-L_0\sigma_{xy}T$. The correction in (b) leaves the data of the three better conducting samples practically unchanged up to $\sim 60\,$K, and for \bsn\ this holds in the entire temperature range.}
	\label{FIG.3.}
\end{figure}

The central result of this study concerns the large low-temperature THE presenting a low-temperature peak of $\kappa_{xy}$, which is orders of magnitude larger than what could be conventionally explained to arise from mobile charge carriers; see Fig.~\ref{FIG.2.}. This low-temperature peak is a common feature of all four samples despite pronounced differences concerning their electrical (Hall) conductivities, which further supports that the underlying mechanism of the low-temperature thermal Hall effect is not related to the mobile charge carriers. A natural candidate for a common mechanism is the almost identical phonon heat conductivity and in fact the peak of $\kappa_{xy}$ at 10--14\,K is located close to the corresponding peak of $\kappa_{xx}$ occurring at 8--12\,K; see Figs.~\ref{FIG.1.}(b,e). A comparison of the temperature dependencies of  $\kappa_{xx}(T)$ and $\kappa_{xy}(T)$ for all four samples is shown in Fig.~\ref{FIG.app_kxx_kxy}(e--h) in the Appendix.

Analogous peak coincidences of $\kappa_{xx}$ and $\kappa_{xy}$ have been reported for various insulating materials over the last years and are discussed as characteristic indicators of a THE that is generated by phonons~\cite{PhysRevB.105.L220301,PhysRevX.12.021025,2022PNAS..11908016C}. A common feature of those materials is a large thermal Hall angle that is expressed by the peak ratios $\left|\kappa_{xy}\right|/\kappa_{xx}$, which are typically in the range of $10^{-3}$ even when the longitudinal $\kappa_{xx}$ values of the different materials differ by up to three orders of magnitude~\cite{PhysRevX.12.021025,PhysRevB.105.115101,2020NatCo..11.5325B,PhysRevLett.124.105901,PhysRevB.99.085136,2022PNAS..11908016C}.

As shown in Fig.~\ref{FIG.3.}, the low-temperature ratio $\kappa_{xy}/\kappa_{xx}$ of each of the four compensated TIs is close to $-10^{-3}$ and varies rather weakly in the temperature range below $\sim 50\,$K where the individual peaks of $\kappa_{xy}$ and $\kappa_{xx}$ evolve. Thus, the pronounced THEs in these TI materials fulfill the above criteria that indicate a phononic origin. In this context, it is also important to mention that these TI samples are nonmagnetic in nature, which rules out the possibility of a low-temperature THE originating from magnons. However, what kind of phononic mechanisms could be capable of generating a THE? According to a recent theoretical proposal~\cite{PhysRevB.105.L220301}, the presence of charged defects in insulators can induce a significant skew scattering of phonons that is linear in magnetic field as a result of an interference between Rayleigh and Lorentz scattering terms and should be observed in the temperature range around and above the phonon maximum of $\kappa_{xx}(T)$. Considering oxide-based insulators, the authors estimate that oxygen vacancies can induce a chirality $\kappa_{xy}/\kappa_{xx}$ of phonon transport of order $-10^{-3}$ that agrees well with our experimental results on the TI materials, but it is also pointed out that the magnitude and even the sign of the expected chirality depend on the relative strength of elastic and attenuation constants~\cite{PhysRevB.105.L220301}. Thus, the above agreement should not be overinterpreted, but the proposed mechanism of a field-linear skew scattering of phonons resulting from charged defects remains a possible source to explain the phonon-related large THE in these TI materials. The occurrence of charged defects in these TIs is naturally expected from the charge compensation that works in such a way that additional acceptors bind the extra electrons coming from donors, which results in negatively charged occupied acceptors and positively charged empty donors. In order to reach insulating bulk behavior, donor and acceptor densities need to be adjusted properly~\cite{PhysRevB.84.165311} and a particular feature of these charge-compensated TI materials is the occurrence of self-organized charge puddles in the low-temperature range~\cite{PhysRevB.93.245149}. The size and connectivity of charge puddles can drastically change the DC electrical resistance and cause characteristic frequency dependencies of the AC conductivities in the microwave to infrared range~\cite{2017NatCo...815545B,PhysRevB.99.161121}. As such effects also influence the screening of charged defects, they can also be relevant for the proposed skew-scattering mechanism of phonons to induce a  large THE in insulators. 

\section{Summary}

To conclude, we have conducted a comparative study of thermal and charge transport in various single crystals of charge compensated TIs. Our key result concerns the field-linear thermal Hall conductivity $\kappa_{xy}$ in all these samples, which becomes most pronounced below about 60~K and peaks around 10~K, where the longitudinal  $\kappa_{xx}(T)$ shows the typical maximum of phonon-dominated heat transport. The corresponding thermal Hall ratios are extraordinarily large $\kappa_{xy}/\kappa_{xx} \sim -10^{-3}$ and vary little between the samples and over the temperature range where the TIs remain insulating. In this respect, the compensated TIs behave similarly to various ionic insulators for which a large THE of phononic origin has been reported~\cite{PhysRevX.12.021025,PhysRevB.105.115101,2020NatCo..11.5325B,PhysRevLett.124.105901,PhysRevB.99.085136,2022PNAS..11908016C}. A skew scattering of phonons induced by charged defects has been suggested as the underlying mechanism of a phononic THE in those ionic insulators~\cite{PhysRevB.105.L220301}, and a similar mechanism could be relevant in these TIs, because the insulating bulk behavior requires charge compensation that naturally induces charged defects via occupied acceptors and empty donors.

The data of this article are available from Zenodo~\cite{sharma_2024_10647723}.

\begin{acknowledgments}
	We thank A. Rosch, J. Hemberger and G. Grissonnanche for helpful discussions. We acknowledge support by the German Research Foundation DFG via Project No.277146847-CRC1238 (Subprojects A04 and B01).
\end{acknowledgments}

\begin{appendix}
	
		
		\section{Electrical Hall versus thermal Hall effect in topological insulators}
		

		\begin{table*}
			\centering
			\caption{Charge-carrier densities of the different samples at 10, 50, and 100\,K which are obtained from the respective low-field slope of the Hall resistivity $\rho_{yx}(B)$ via $n= B/(e\, \rho_{yx}(B))$. The last four columns list the longitudinal $\kappa_{xx}$ and transverse $\kappa_{xy}$ thermal conductivities measured in the maximum field of 14\,T at temperatures of 10 and 50\,K.}
			\addtolength{\tabcolsep}{7pt}   
			\begin{tabular}{l c c c c c c c }
				\hline
				\hline\\
				Sample & n$^{10 K}$& n$^{50 K}$& n$^{100 K}$& $\kappa_{xx}$$^{10 K}$& $\kappa_{xy}$$^{10 K}$& $\kappa_{xx}$$^{50 K}$& $\kappa_{xy}$$^{50 K}$\\
				& (cm$^{-3}$)& (cm$^{-3}$)& (cm$^{-3}$)& (W/Km)& (mW/Km) & (W/Km)& (mW/Km)\\			
				\hline \\
				BiSbTeSe$_2-$S1&$+2.9\times10^{17}$&$+8.9\times10^{18}$&$+7.3\times10^{17}$&4.86&-7.4& 1.89&-0.7\\ 
				BiSbTeSe$_2-$S2&$-4.2\times10^{16}$&$-1.3\times10^{16}$&$+2.7\times10^{17}$&4.18&-6.6& 1.92&-0.8\\ 
				Bi$_{1.5}$Sb$_{0.5}$Te$_{1.7}$Se$_{1.3}$&$-6.8\times10^{16}$&$-1.5\times10^{17}$&$-2.0\times10^{17}$&4.64&-3.9& 1.93&-1.6\\ 
				Bi$_{1.08}$Sn$_{0.02}$Sb$_{0.9}$Te$_{2}$S&$-7.9\times10^{14}$&$-4.9\times10^{14}$&$-5.7\times10^{14}$&6.06&-7.5& 2.68&-3.5\\
				\hline\\
			\end{tabular}
			\label{Table.1}
		\end{table*}
		
		Table~\ref{Table.1} lists the charge-carrier densities derived from the low-field Hall resistivity data of the studied samples for some characteristic temperatures, and also gives the values of the obtained longitudinal and transverse thermal conductivities in a field of 14~T.		
		
\begin{figure*}[t]
			\centering
			\hspace*{-0.8cm}
			\includegraphics[scale=0.55]{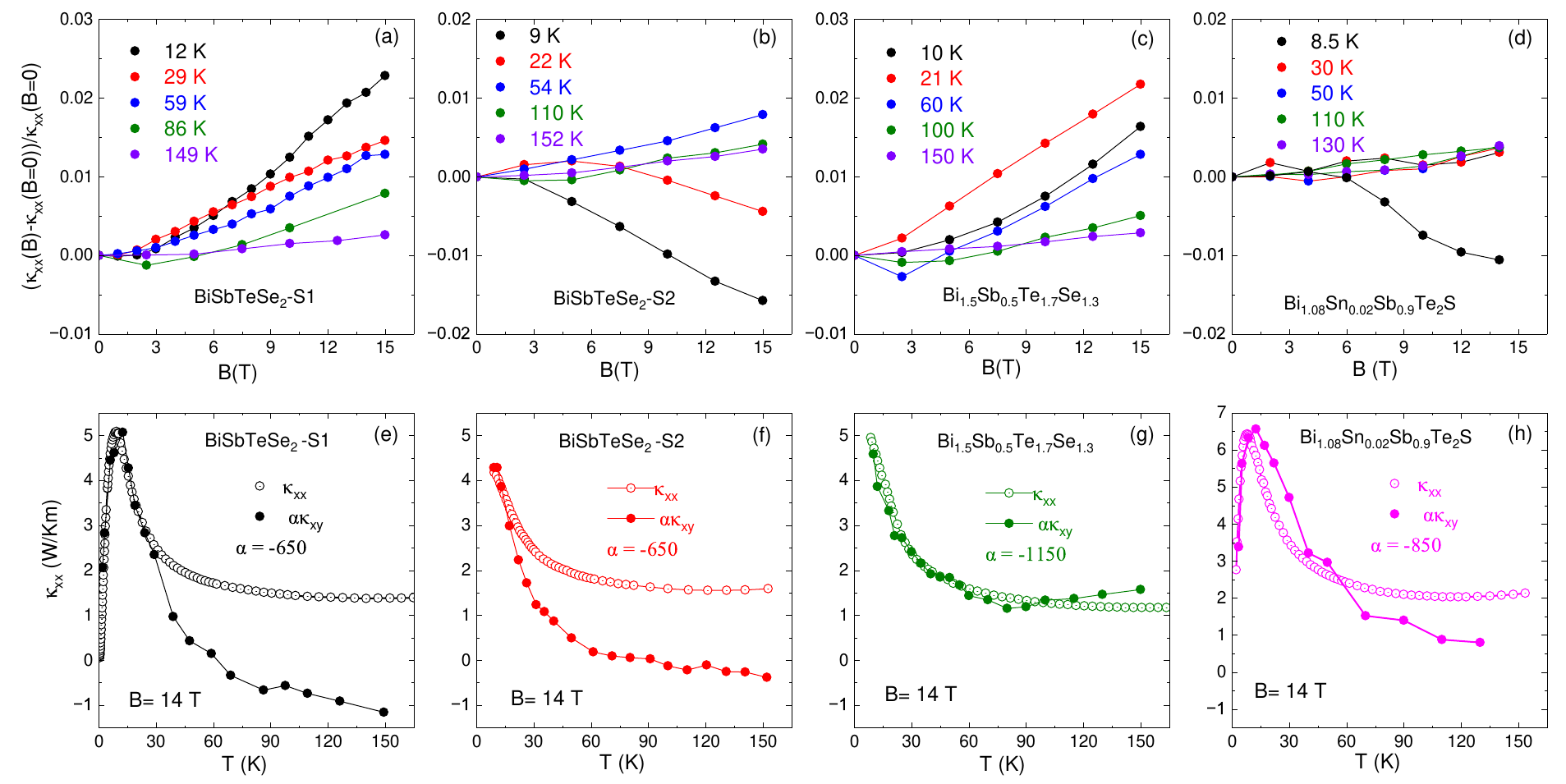}
			\caption{(a--d) Magnetic field dependence of $\kappa_{xx}$ for all charge compensated TI samples that remains in the range of 2~\% at all temperatures. (e--h) Comparison of the temperature dependencies of the longitudinal  and  transverse thermal conductivities by the rescaling $\kappa_{xy}$ measured in a magnetic 	
				field of 14\,T with a factor $\alpha$ such that it meets the corresponding $\kappa_{xx}$ around the low-temperature maximum.}
			\label{FIG.app_kxx_kxy}
		\end{figure*}

		Figure~\ref{FIG.app_kxx_kxy}(a--d) presents the longitudinal thermal conductivity $\kappa_{xx}$ as a function of magnetic field for different termperatures.  In all four samples, this magnetic-field dependences is very weak as the relative variation remains below about 2\,\% over the entire temperature range. Panels~(e--h) of Fig.~\ref{FIG.app_kxx_kxy} show that the thermal Hall conductivities $\kappa_{xy}$ are strongly temperature dependent and follow a similar low-temperature dependence as the phonon-dominated longitudinal thermal conductivity $\kappa_{xx}$. This kind of scaling behavior between the field-linear $\kappa_{xy}$ and the almost field-independent $\kappa_{xx}$ is a typical indication of a phononic origin of the thermal Hall effect.

		
\begin{figure*}
			\centering
			\hspace*{-0.8cm}
			\includegraphics[scale=0.48]{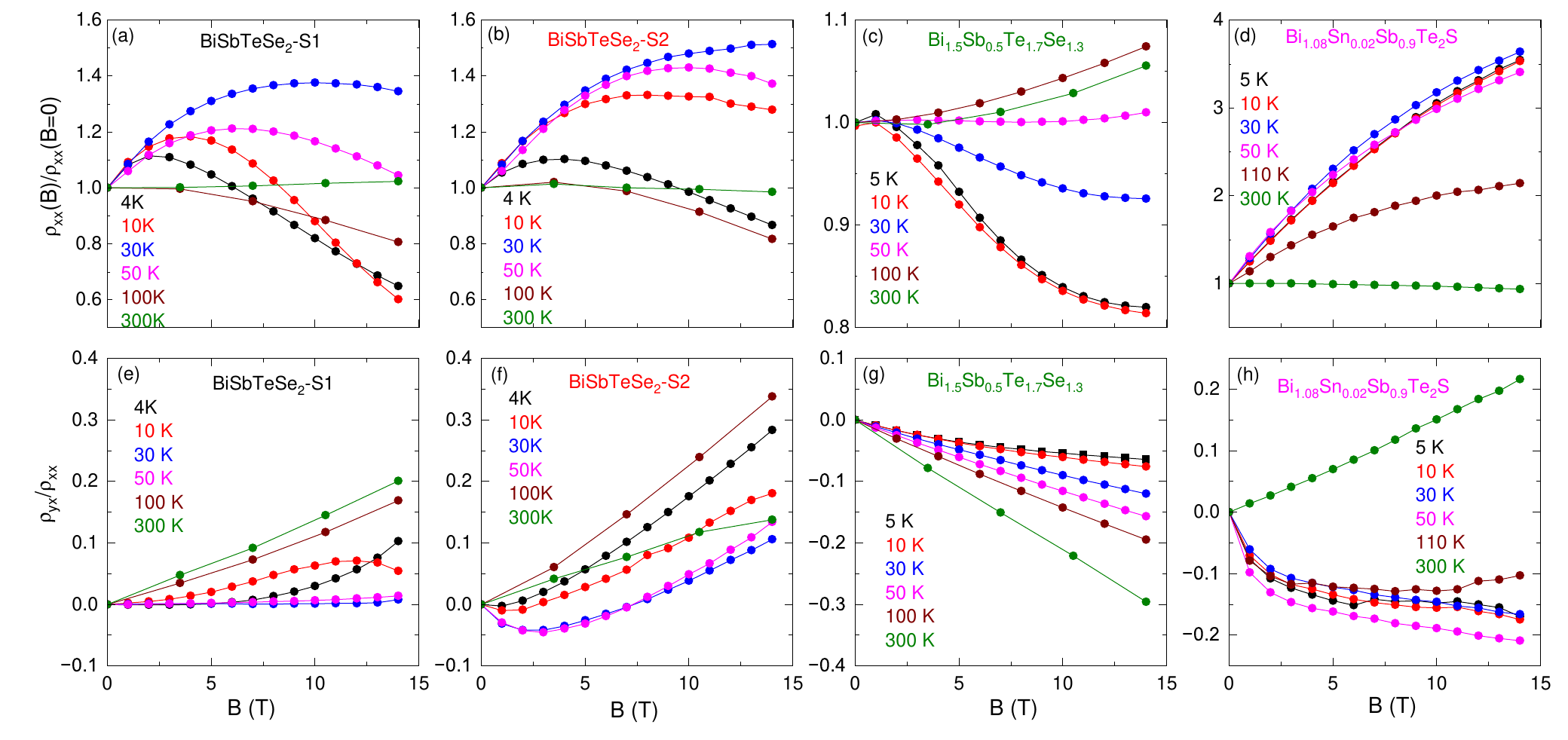} 
			\caption{(a--d) Normalized resistivity $\rho_{xx}(B)$/$\rho_{xx}(B=0)$ and (e--h) Hall ratio $\rho_{yx}(B)/\rho_{xx}(B)$ of charge-compensated TIs.}
			\label{FIG.app_rhoxx_rhoyx}
		\end{figure*}

		Figure~\ref{FIG.app_rhoxx_rhoyx} gives an overview of the magnetoresistance and the electrical Hall ratio of the four compensated TIs. At stabilized temperatures and constant magnetic fields, the Hall resistivity $\rho_{yx}(B)$ and the longitudinal resistivity $\rho_{xx}(B)$ were obtained by, respectively, anti-symmetrizing and symmetrizing the corresponding transverse ($U_y$) and longitudinal ($U_x$) voltages measured in the field range up to $\pm 14\,$T. As seen in panels~(a--d), the normalized magnetoresistance is large in all samples ranging from several $10\,\%$ in \bsts\ and \bsb\ to a factor of 3 in \bsn . The Hall ratios, shown in (e--f), reach values up about $30\,\%$, but are of different signs depending on the sample and/or the temperature. Generally, the Hall ratios show non-linear field dependencies at low temperatures and approach field-linear behavior towards higher temperature. 
		
\begin{figure*}
	\centering
	\hspace*{-0.8cm}
	\includegraphics[scale=0.55]{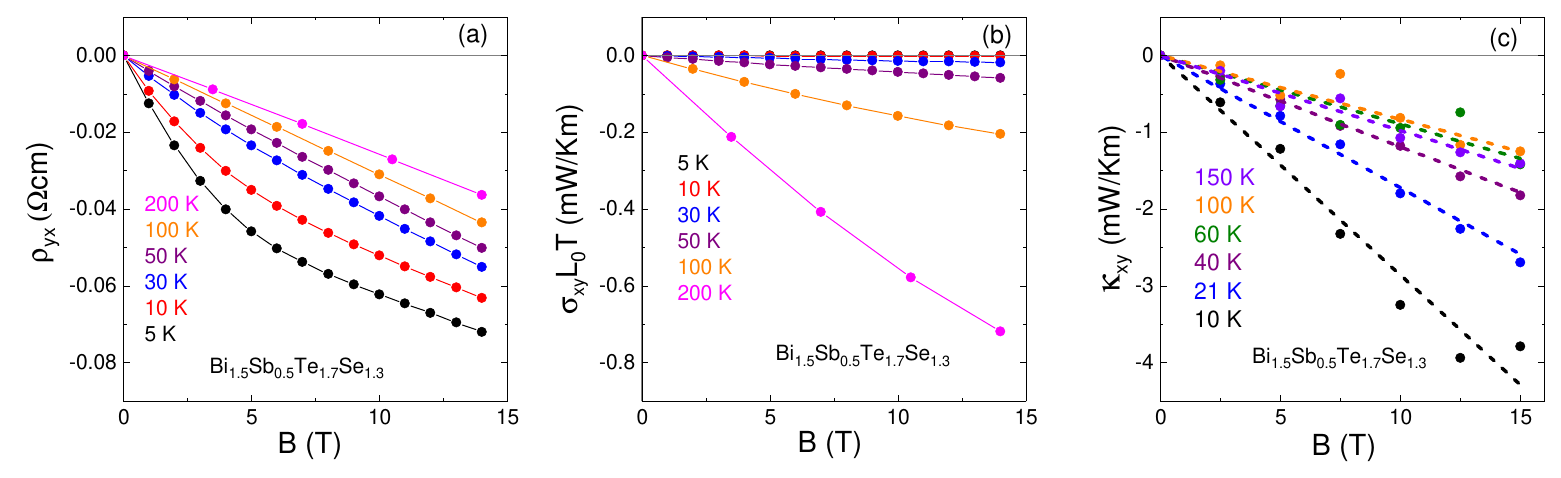}
	\caption{(a) Hall resistivity $\rho_{yx}(B)$ and (b) the corresponding electronic contribution to the thermal Hall conductivity $\kappa_{xy}^{el}(B)$ in comparison to (c) the measured total Hall conductivity $\kappa_{xy}(B)$ of \bsb . The dashed lines in (c) are linear fits that were used to calculate the temperature dependent $\kappa_{xy}(T)$ at constant fields (see text).}
	\label{FIG.app_WFL_bsb}
\end{figure*}

\begin{figure*}
	\centering
	\hspace*{-0.8cm}
	\includegraphics[scale=1.37]{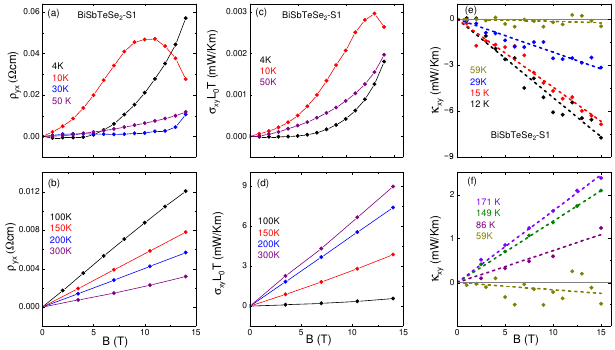}
	\caption{(a,b) Hall resistivity $\rho_{yx}(B)$ and (c,d) the corresponding electronic contribution to the thermal Hall conductivity $\kappa_{xy}^{el}(B)$ in comparison to (e,f) the measured total Hall conductivity $\kappa_{xy}(B)$ of \bsts\ (sample 1). The dashed lines in (e,f) are linear fits that were used to calculate the temperature dependent $\kappa_{xy}(T)$ at constant fields (see text).}
	\label{FIG.app_WFL_bsts-S1}
\end{figure*}

\begin{figure*}
	\centering
	\hspace*{-0.8cm}
	\includegraphics[scale=0.55]{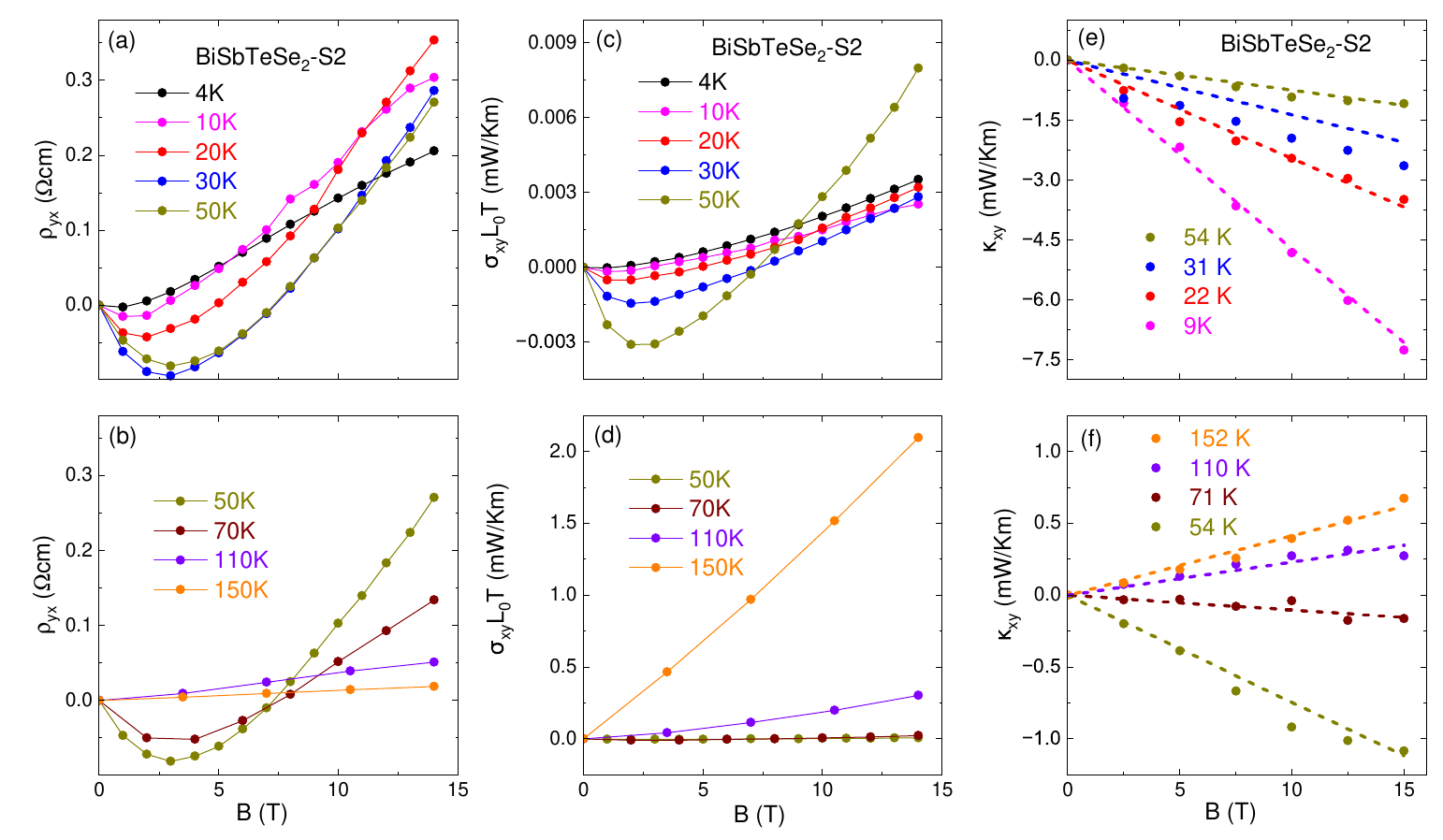} 
	\caption{(a,b) Hall resistivity $\rho_{yx}(B)$ and (c,d) the corresponding electronic contribution to the thermal Hall conductivity $\kappa_{xy}^{el}(B)$ in comparison to (e,f) the measured total Hall conductivity $\kappa_{xy}(B)$ of \bsts\ (sample 2). The dashed lines in (e,f) are linear fits that were used to calculate the temperature dependent $\kappa_{xy}(T)$ at constant fields (see text).}
	\label{FIG.app_WFL_bsts-S2}
\end{figure*}

\begin{figure*}
	\centering
	\hspace*{-0.8cm}
	\includegraphics[scale=0.55]{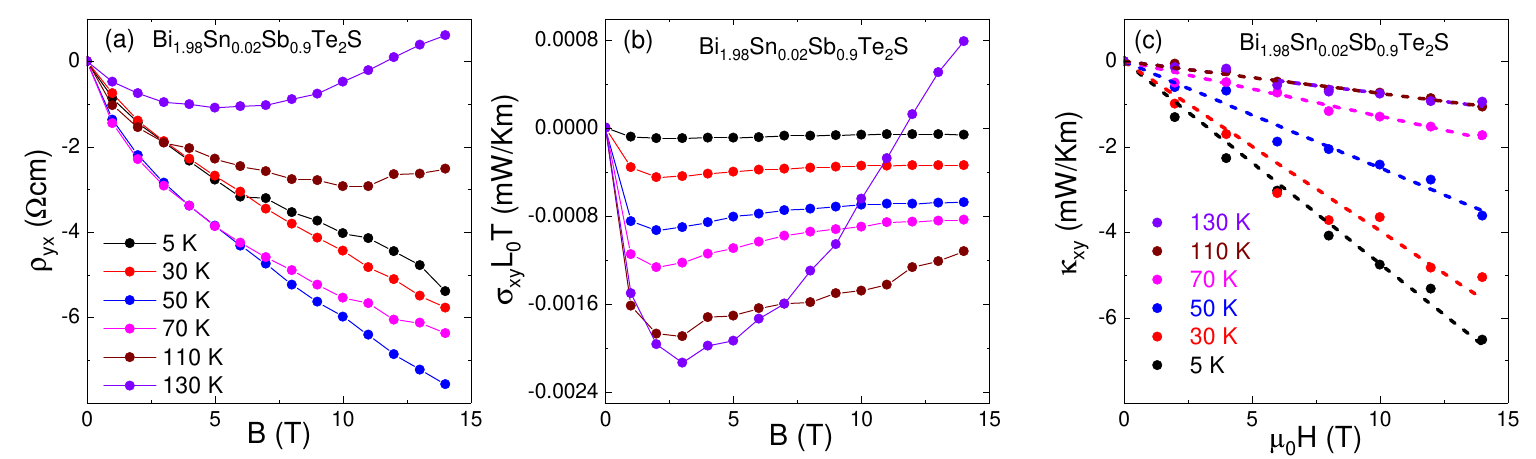}
	\caption{(a) Hall resistivity $\rho_{yx}(B)$ and (b) the corresponding electronic contribution to the thermal Hall conductivity $\kappa_{xy}^{el}(B)$ in comparison to (c) the measured total Hall conductivity $\kappa_{xy}(B)$ of \bsn . The dashed lines in (c) are linear fits that were used to calculate the temperature dependent $\kappa_{xy}(T)$ at constant fields (see text).}
	\label{FIG.app_WFL_bsn}
\end{figure*}

		From $\rho_{xx}(B)$ and $\rho_{yx}(B)$ the electrical Hall conductivity $\sigma_{xy}(B)=\rho_{yx}/(\rho_{xx}^2+\rho_{yx}^2)$ is calculated, from which we estimate the electronic contribution to the thermal Hall conductivity $\kappa_{xy}^{el}(B) = L_0\,\sigma_{xy}(B)\,T$ by using the  Wiedemann-Franz law with temperature $T$ and Lorenz number L$_0 \approx 2.44\times10^{-8}$ V$^2$K$^{-2}$. In Figs.~\ref{FIG.app_WFL_bsb}--\ref{FIG.app_WFL_bsn}, the Hall resistivities $\rho_{yx}(B)$ are shown in the left panels while the obtained $\kappa_{xy}^{el}(B)$ are displayed in the middle panels and can be compared to the measured total thermal Hall conductivities $\kappa_{xy}(B)$ shown in the right panels. As discussed above, in all four samples the longitudinal thermal conductivities $\kappa_{xx}$ vary by less than 2\% up to a magnetic field of 14~T, see Fig.~\ref{FIG.1.}(c) and~\ref{FIG.app_kxx_kxy}(a--d), whereas the thermal Hall conductivities $\kappa_{xy}$ remain field-linear over the entire temperature range, see Fig.~\ref{FIG.1.}(f) and~Figs.~\ref{FIG.app_WFL_bsb}---\ref{FIG.app_WFL_bsn} (right panels).

	\end{appendix}



\providecommand{\noopsort}[1]{}\providecommand{\singleletter}[1]{#1}%

\end{document}